\documentclass[12pt,fleqn]{article}
\usepackage{graphicx}
\usepackage{cite,epsfig,color}

\def\be{\begin{equation}}
\def\ee{\end{equation}}
\def\bea{\begin{eqnarray}}
\def\eea{\end{eqnarray}}

\def\bbuildrel#1_#2^#3{\mathrel{\mathop{\kern 0pt#1}\limits_{#2}^{#3}}}
\def\slash#1{\setbox0=\hbox{$#1$}#1\hskip-\wd0\dimen0=5pt\advance
       \dimen0 by-\ht0\advance\dimen0 by\dp0\lower0.5\dimen0\hbox
         to\wd0{\hss\sl/\/\hss}}
\def\gev{{\rm GeV}}

\def \ap{A_{\|}}
\def \app{{A}_{\bot}}
\def \kstar{{K^*}}
\def \trans{{T}}

\def \Re{{\rm {Re}}}

\def \braket#1#2#3{\langle #1|#2| #3\rangle}

\def \ea{{\it et al.}}
\def \eq#1{Eq.~(\ref{#1})}

\def \nnu{\nonumber}

\def \ol#1{\overline{#1}}

\def \rf{Ref.~\cite}

\def \cseff{C_7^{\rm {eff}}}
\def \cseffP{{C_7^{\rm {eff}}}^\prime}
\def \ceff{C_9^{\rm {eff}}}

\def \cten{C_{10}}

\def \a{\alpha}

\def \g{\gamma}

\def \m{\mu}

\newcommand{\gae}{\lower 2pt \hbox{$\, \buildrel {\scriptstyle >}\over {\scriptstyle
\sim}\,$}}
\newcommand{\lae}{\lower 2pt \hbox{$\, \buildrel {\scriptstyle <}\over {\scriptstyle
\sim}\,$}}

\newcommand{\spp}{\vphantom{\bigg(}}
\begin{document}

\begin{titlepage}

\begin{flushright}
FERMILAB-PUB-06-457-T\\
CERN-PH-TH/2006-255\\
UAB-FT-617\\
hep-ph/0612166\\[2cm]
\end{flushright}

\begin{center}
\setlength {\baselineskip}{0.2in} {\bf \large Huge right-handed
current effects in $B\to K^* ( K \pi) \ell^+ \ell^-$ in \\[0.2cm]
supersymmetry}
\\[2cm]

\setlength {\baselineskip}{0.2in} {\large Enrico
Lunghi$^1$, Joaquim Matias$^{2}$\footnote{On leave of absence at CERN-TH, CH-1211 Geneva 23}}\\[5mm]

$^1$~{\it
 Fermi National Accelerator Laboratory \\
 P.O. Box 500, Batavia, IL 60510-0500, USA\\
 E-mail: {\rm lunghi@fnal.gov}
}\\[5mm]
$^2$~{\it
 IFAE, Universitat Aut\`onoma de Barcelona \\
08193 Bellaterra, Barcelona, Spain \\
E-mail: {\rm matias@ecm.ub.es}
}\\[3cm]

{\bf Abstract}\\[5mm]
\end{center}
\setlength{\baselineskip}{0.2in}
Transverse asymmetries in the decay $B\to K^*(K \pi) \ell^+
\ell^-$ are an extremely sensitive probe of right-handed
flavour-changing neutral currents. We show how to include the
contribution from the chiral partner of the electromagnetic
operator on the transverse asymmetries at NLO in QCD
factorization. We then consider supersymmetric models with
non-minimal flavour violation in the down-squark sector.
 We include all the relevant experimental
constraints and present a numerical formula for $B\to X_s \gamma$
that takes into account the most recent NNLO calculations. We show
that the flavour-changing parameters of these models are poorly
constrained by present data and allow for large effects on the
transverse asymmetries that we consider.

\end{titlepage}

\section{Introduction}
B decays offer a unique opportunity to explore the flavour
structure of the theory lying beyond the Standard
Model~\cite{fs,fs1,fs2,fs3,fs4}. Given the scenario depicted by
present data, the search for new physics in the flavour sector is
evolving more and more toward precision analyses. On the one
side, new methods are being developed to produce more accurate
predictions focusing, in particular, on the problem of large
$\Lambda/m_b$ corrections (see for instance Ref.~\cite{dmv1} in
the context of $B \to \pi\pi$ decays and Ref.~\cite{dmv2} for $B_s
\to KK$ decays). On the other side, new observables are being
proposed to test specific types of new physics (presence of
right-handed
currents~\cite{Melikhov:1998cd,CS:etal,CS:etal1,Kruger:2005ep},
isospin breaking beyond the SM~\cite{fm,fm1}, etc.). In
particular, one of the most important targets of present searches
is to find observables that can test the chiral structure of the
fundamental theory  lying beyond the SM.

In a previous paper~\cite{Kruger:2005ep} a set of observables
based on the angular distribution of the decay $B\to K^*(K \pi)
\ell^+ \ell^-$ were analysed at NLO in the SM in the framework of
QCD~Factorization. They provide information of the $K^*$ spin
amplitudes~\cite{FK:etal,CS:etal,CS:etal1} that are useful to
search for right-handed currents. The goal was to identify the
most robust observables, i.e., those less affected by hadronic
uncertainties, to search for new physics originated by
right-handed currents. The most promising observables were found
to be the transverse asymmetries (see Ref.~\cite{Kruger:2005ep})
due to the exact cancellation, at leading order, of the poorly
known soft form factors. This cancellation is basically not
spoiled when including NLO corrections.

In Ref.~\cite{Kruger:2005ep} a model-independent analysis was done
to test the possible impact of right-handed currents on those
promising observables. It remained to be explored whether a
well-motivated model, once all kinds of constraints are included,
can still naturally lead to large deviations.

There are several models that can produce right-handed currents,
such as, left-right-symmetric models with or without spontaneous
CP violation. However, some of these models have already been
ruled out (see for instance Ref.~\cite{ball}). The aim of the
present paper is to show the possible impact that a well-motivated
Minimal Supersymmetric Model (MSSM), with non-minimal flavour
changing and R-parity conservation, has on the transverse
asymmetries and the polarization fraction (see \cite{fulvia} for
universal extra-dimensions case). The corresponding integrated
observables are analysed as well. Because of its experimental
interest~\cite{Aubert:2006vb}, we have included also the
prediction for the longitudinal fraction of $K^*$ polarization,
even though it was already shown in~\cite{Kruger:2005ep} that it
is a difficult task to extract clean information concerning new
physics out of this observable. Apart from uncertainties coming
from soft form factors, another important source of theoretical
error comes from the uncontrolled $\Lambda/m_b$ contributions. In
order to explore the impact of these unknown corrections, we allow
for a $\pm 10\%$ error (this comes from taking $\Lambda$ in a
range between 200 and 400 MeV, $m_b$ of the order of 5 GeV, and
assuming all coefficients of order 1) on each individual
amplitude.

The structure of the paper is the following. In
Section~\ref{sec:asymmetries}, we generalize the transverse amplitudes
at NLO in order to incorporate the extra contribution coming from the
chiral partner of the electromagnetic operator and we define the
observables.  In Section~\ref{sec:gluino} we describe the structure of
the squark mass matrix to show how the sources of flavour changing
enter. We will focus on the dominant contribution coming from penguin
diagrams involving a gluino and down squarks, and we will use the
complete result for the Wilson coefficients
following~\cite{Bobeth:1999ww} that generalizes the results given in
the approximate formulae~\cite{Lunghi:1999uk,kim}. We will chose a
minimal set of free parameters (gluino and down squark masses together
with only one mass insertion), which are sufficient to illustrate the
large impact on the interesting observables. Finally we impose in
Section~\ref{sec:constraints} all relevant experimental constraints,
including the very last results for $B\to X_s
\gamma$~\cite{Misiak:2006zs,Misiak:2006ab,Lee:2006wn,Becher:2006pu}.
Next, we describe in Section~\ref{sec:results} the numerical results
focusing on some representative cases to show the huge impact that the
model has on these observables. We also show the constraint on the
squark and gluino masses implied by a measurement of a large effect on
the transverse asymmetries. We conclude in
Section~\ref{sec:conclusions}.

\section{The transverse asymmetries}
\label{sec:asymmetries}
The effective Hamiltonian describing the quark transition $b\to s
l^+l^-$~\cite{Melikhov:1998cd,FK:etal,CS:etal,CS:etal1} is given by
\cite{wilson:coeffs:SM,wilson:coeffs:SM1,wilson:coeffs:SM2,wilson:coeffs:SM3}:
\be\label{heff}
{\mathcal{H}}_{\rm eff}=-\frac{4 G_F}{\sqrt{2}} V_{tb}^{} V_{ts}^*
\sum_{i=1}^{10}
[C_i (\mu) {\mathcal{O}}_{i}(\mu)  + C_i^\prime (\mu) {\mathcal{O}}_{i}^\prime
(\mu)],
\ee
where in addition to the SM operators we have added the chirally
flipped partners.  For the complete set of operators
(${\mathcal{O}}_{i}^{(\prime)}(\mu)$) and Wilson coefficients
($C_i^{(\prime)}(\mu)$) in the SM and beyond, we refer the reader to
\cite{wilson:coeffs:SM,wilson:coeffs:SM1,wilson:coeffs:SM2,
wilson:coeffs:SM3,ball,LRmodel,LRmodel1,Borzumati:1999qt,Borzumati:1999qt1,Borzumati:1999qt2}.
In what follows we will use the same conventions as
in~\cite{Kruger:2005ep}.

We will be specially interested here in the two electromagnetic
partner operators:
\bea\label{operator:basis}
{\mathcal{O}}_{7} = \frac{e}{16\pi^2} m_b
(\bar{s} \sigma_{\mu \nu} P_R b) F^{\mu \nu}, \quad
{\mathcal{O}}_{7}^\prime = \frac{e}{16\pi^2} m_b
(\bar{s} \sigma_{\mu \nu} P_L b) F^{\mu \nu}
\eea
and in the semileptonic operators:
\bea
{\mathcal{O}}_{9} = \frac{e^2}{16\pi^2}
(\bar{s} \gamma_{\mu} P_L b)(\bar{l} \gamma^\mu l), \quad
{\mathcal{O}}_{10}=\frac{e^2}{16\pi^2}
(\bar{s}  \gamma_{\mu} P_L b)(  \bar{l} \gamma^\mu \gamma_5 l),
\eea
where $P_{L,R}= (1\mp \g_5)/2$ and $m_b \equiv m_b(\mu)$ is the
running mass in the $\ol{\mbox{MS}}$ scheme.  From the effective
Hamiltonian it is straightforward to compute the matrix element for
the decay $B \rightarrow K^*(\rightarrow K\pi) l^+l^-$:
\bea
\label{matrix:ele}
{\mathcal M}
&=&
\frac{G_F\a}{\sqrt{2}\pi}V_{tb}^{}V_{ts}^* \; \bigg\{
\nnu \\ &&
\bigg[\ceff\braket{K \pi}{(\bar{s}\g^{\mu}P_Lb)}{B} -\frac{2m_b}{q^2}
\braket{K\pi}{\bar{s}i\sigma^{\mu\nu}q_{\nu}(\cseff P_R+\cseffP  P_L)b}{B}\bigg]
(\bar{l}\g_{\m}l)
\nnu\\ &&
+ \; \cten\braket{K\pi}{(\bar{s}\g^{\mu}P_Lb)}{B}(\bar{l}\g_{\mu}\g_5 l)\bigg \},
\eea
where $q$ is the four-momentum of the lepton pair. The explicit
form of the four hadronic matrix elements can be found
in~\cite{Kruger:2005ep}.

Our goal in this section will be to generalize the formulae
in~\cite{Kruger:2005ep} to describe the angular distribution and
transversity amplitudes at NLO, in the presence of the chirally
flipped operator ${\cal O}^{\prime}_7$. The
transversity amplitudes corresponding to the four physical $K^*$
spin amplitudes  $A_{\bot}$, $A_{\|}$, $A_{0}$ and $A_t$ are
related to the helicity amplitudes, also used in literature
through: \be\label{hel:trans} A_{\bot,\|} = (H_{+1}\mp
H_{-1})/\sqrt{2}, \quad A_0=H_0, \quad A_t=H_t. \ee

Each spin amplitude splits in a left-handed and a right-handed
component and in our observables discussed below we introduce the
shorthand notation:
\bea A_i A^*_j\equiv A^{}_{i L}(s) A^*_{jL}(s)+ A^{}_{iR}(s)
A^*_{jR}(s) \quad (i,j  = 0, \|, \perp). \eea

The generalization to include the impact of the dipole operator
$O_7^\prime$  is achieved by taking the transversity amplitudes
(we focus here on $A_{\bot L,R}$, $A_{\| L,R}$ and $A_{0L,R}$)
obtained from the above matrix element\footnote{For an analysis of form 
factors in the context of QCD light-cone sumrules see \cite{pr}}:
\be\label{a_perp}
A_{\bot L,R}=N \sqrt{2} \lambda^{1/2}\bigg[
(\ceff\mp\cten)\frac{V(s)}{m_B +m_\kstar}+\frac{2m_b}{s} (\cseff + \cseffP)
T_1(s)\bigg],
\ee
\be\label{a_par}
A_{\| L,R}= - N \sqrt{2}(m_B^2- m_\kstar^2)\bigg[(\ceff\mp \cten)
\frac{A_1 (s)}{m_B-m_\kstar}
+\frac{2 m_b}{s} (\cseff - \cseffP) T_2(s)\bigg],
\ee
\bea\label{a_long}
A_{0L,R}&=&-\frac{N}{2m_\kstar\sqrt{s}}\bigg[
(\ceff\mp \cten)\bigg\{(m_B^2-m_\kstar^2 -s)(m_B+m_\kstar)A_1(s)
-\lambda \frac{A_2(s)}{m_B +m_\kstar}\bigg\}\nnu\\
&+&{2m_b}(\cseff - \cseffP) \bigg\{
 (m_B^2+3m_\kstar^2 -s)T_2(s)
-\frac{\lambda}{m_B^2-m_\kstar^2} T_3(s)\bigg\}\bigg],
\eea
where $\lambda$ and $N$ are defined as in \cite{Kruger:2005ep},
 and perform the
following substitutions:
\be\label{subs:Tis}
(\cseff+\cseffP)
 T_i \rightarrow {\cal T}_i^+, \quad
(\cseff-\cseffP)
 T_i \rightarrow {\cal T}_i^-,
\quad
\ceff \rightarrow C_9\quad (i=1,2,3),
\ee
with the Wilson coefficients $C_{9,10}$ taken at NNLL order (in the
sense of \rf{Beneke:2001at}). The ${\cal T}_{i}^{\pm}$ in
\eq{subs:Tis} are given by
\bea
{\cal T}_1^{\pm}={\cal T}^{\pm}_\perp, \quad {\cal T}_2^{-}=\frac{2
E_{\kstar}}{m_B}{\cal T}_\perp^{-},
 \quad {\cal T}_3^{-}={\cal T}_\perp^{-}+{\cal T}_\parallel^{-}.
\eea
${\cal T}_a^{\pm}$ ($a=\bot,\|$) contain factorizable (f) and
non-factorizable (nf) contributions \cite{Beneke:2001at} and it is
defined by:
\bea\label{nlodef}
{\cal T}_\perp^{\pm} &=& \xi_\perp(0)
\Bigg\{ C_\perp^{(0,\pm)} \frac{1}{(1-s/m_B^2)^2} +
\frac{\alpha_s}{3\pi} \Bigg[
\frac{C_\perp^{(1,\pm)}}{(1-s/m_B^2)^2}+ \kappa_\perp
  \lambda^{-1}_{B,+}\int_0^1 du\,\Phi_{K^*, \perp}(u)\nnu\\
&\times& [T_{\perp,+}^{(\mathrm{f}\pm)}(u)+
T_{\perp,+}^{(\mathrm{nf}\pm)}(u)] \Bigg] \Bigg\},
\eea
where the symbol $\pm$ stands for the substitution of $\cseff \to
\cseff+\cseffP$ (for $+$) and $\cseff \to \cseff-\cseffP$ (for
$-$), wherever $\cseff$ appears.  ${\cal T}_{\|}^{-}$ is defined
in a completely analogous way from the definition of ${\cal
T}_{\|}$ \cite{Beneke:2001at}; however, in this case, the
substitution is always $\cseff \to \cseff-\cseffP$. The Wilson
coefficients of the standard basis ${\cal O}_i$ run at NLO,
following~\cite{Bobeth:1999mk}, while the running of the chirally
flipped $\cseffP$ is done at LO following~\cite{Cho:1993zb}.

We can now insert the complete transverse amplitudes into our set of
observables:
\be\label{def:asymmetries}
A^{(1)}_{\trans}({s})=\frac{-2\Re(\ap^{}\app^*)}{|\app|^2 +
|\ap|^2},\quad A^{(2)}_{\trans}({s})=\frac{|\app|^2 -
|\ap|^2}{|\app|^2 + |\ap|^2}, \quad F_L(s) = \frac{|{A}_0|^2}{|{
A}_0|^2 + |{A}_{\|}|^2 + |A_\perp|^2}, \ee
corresponding to the transverse asymmetries and the $K^*$ polarization
fraction.  The reason to choose this small subset of observables is
due to the robustness of $A^{(1,2)}_{\trans}({s})$ in front of the NLO
corrections, as it was found in~\cite{Kruger:2005ep}, and to the
experimental interest of $F_L(s)$. However, we will show here that it
is difficult to extract clean information concerning new physics from
this observable $F_L(s)$. We will also consider the corresponding
integrated quantities: ${\mathcal A}^{(1)}_{\trans}$, ${\mathcal
A}^{(2)}_{\trans}$ obtained integrating numerator and denominator of
the corresponding observables over the low-$s$ region $1\; {\rm GeV^2}
< s < 6 \; {\rm GeV^2}$. It is interesting to observe
that working in the helicity basis defined in Eq.  (\ref{hel:trans})
within the SM (where in particular $\cseffP=0$) one recovers the
quark-model prediction $H_{+1}=0$. The physical reason is that the
combination of the $s$ quark produced with an helicity -1/2 by weak
interactions, once combined with the light quark can only form a $K^*$
in an helicity state -1 or 0. This translates approximately in
$A_T^{(1)}\sim 1$ and $A_T^{(2)} \sim 0$ in the SM. Indeed the
different sign contribution of $\cseffP$ in $A_\bot$, as it can be
seen in Eq.(\ref{a_perp}), versus $A_\|$ (Eq. (\ref{a_par})) generates
an interference term proportional to $ 4 {\cseffP}^2
{\hat m_b}^2/{\hat s}^2$ strongly sensitive to the presence of
right-handed currents.

In a previous paper~\cite{Kruger:2005ep} we observed in a
model-independent way that a relatively small contribution to $\cseffP$
has a strong impact on those asymmetries.  On the other hand, in
Ref.~\cite{Kruger:2005ep} it was also shown that the impact of
$C_{9,10}$ is quite small and subleading when compared with $\cseffP$
(due to the 2 ${\hat m_b}/{\hat s}$ factor), once the constraint from
$B\to X_s \ell^+\ell^-$ is taken into account (see Fig.5 and 6 in 
\cite{Kruger:2005ep}). The inclusion of their
chiral partners via the substitutions: $C^{(\mathrm{eff})}_{9,10}\to
C^{(\mathrm{eff})}_{9,10}+ C^{(\mathrm{eff})\prime}_{9,10}$ in
\eq{a_perp}, $C_{9,10}^{(\mathrm{eff})}\to C^{(\mathrm{eff})}_{9,10}-
C^{(\mathrm{eff})\prime}_{9,10}$ in Eqs.~(\ref{a_par}) and
(\ref{a_long}) entering at the same level as ${\cal O}_{9,10}$ will
have a similarly small impact.

For this reason the present analysis will focus on models with large
effects on $\cseffP$ only. In the next section we will check it
explicitly for a well-motivated supersymmetric model, taking into
account all possible constraints.

Another observable that is very sensitive to a non-vanishing $\cseffP$
is the time-dependent CP asymmetry in $B\to K^*
\gamma$~\cite{Atwood:1997zr,Grinstein:2004uu,Atwood2} whose theoretical errors
on the leading power contributions are small.  It is very important to
test both the $q^2=0$ limit with this asymmetry and the whole spectrum
using the asymmetries $A_T^{(1)}(s)$ and $A_T^{(2)}(s)$. However, at
the experimental level at LHCb the time-dependent analysis in $B\to
K^* \gamma$ requires to look at a final state, which is a CP
eigenstate ($K^{*0} \to K^0_S \pi^0$). This is considered very
difficult at LHCb \cite{Ulrik} but possible  at the
B-Factories \cite{exp}.  Despite of this, a measurement of both set of
observables $B\to K^* \gamma$ and $A_T^{(1)}(s)$ and $A_T^{(2)}(s)$
with high precision could be also useful to set bounds on the
sub-leading $O_{9,10}^{(')}$ contributions.

\section{Gluino-mediated FCNC}
\label{sec:gluino}
As an example of a new physics model that allows for large
contributions to $\cseffP$, we consider an R-parity-conserving
MSSM with non-minimal flavour changing in the down-squarks
soft-breaking terms. We define the model at the electroweak scale
and implement the resummation of large-$\tan\beta$
effects~\cite{Babu:1999hn,Hamzaoui:1998nu,Isidori:2001fv,Buras:2002wq,Buras:2002vd,Dedes:2002er}
in the quark mass eigenstate basis~\cite{Foster:2005wb}.  We adopt
the notation and conventions of Ref.~\cite{Foster:2005wb}.

The soft-breaking terms are given in the physical super-CKM basis.
In this basis, rigid superfield rotations are used to diagonalize
the physical quark mass matrices, i.e., this is the basis in
which, after the integration of the soft-breaking terms, the quark
masses and the CKM matrix coincide with the observed ones. The
down-squark mass matrix in the physical SCKM basis is
\bea
{\cal M}^2_{\tilde d} & = &
\pmatrix{
m^2_{d,LL} + F_{d,LL} + D_{d,LL} &
m^2_{d,LR} + F_{d,LR} \cr
\left(m^2_{d,LR} + F_{d,LR}\right)^\dagger &
m^2_{d,RR} + F_{d,RR} + D_{d,RR} \cr
}
\eea
where the $F$-terms are $F_{d,LL} = F_{d,RR}^\dagger =
m_d^{(0)\dagger} m_d^{(0)}$, $F_{d,LR} = -\mu \tan\beta
m_d^{(0)\dagger}$ and the $D$-terms are $D_{d,LL} = m_Z^2 \cos 2
\beta (-1/2 + \sin^2 \theta_W /3)$, $D_{d,RR} = -m_Z^2 \cos 2
\beta \sin^2 \theta_W /3$. In the above formulae,  $m_d^{(0)}$ is
the tree-level down-quark mass matrix: the physical mass matrix is
obtained only after adding the supersymmetric corrections, $m_d =
m_d^{(0)} + \delta m_d = {\rm diag}(m_d,m_s,m_b)$.

We parametrize off-diagonal entries of the down-squark mass matrix
in terms of mass insertions ($A,B = L,R$):
\bea
\left(\delta^d_{AB} \right)_{ij}\equiv \left(m_{d,AB}^2\right)_{ij} /
\sqrt{ \left(m^2_{d,AA}\right)_{ij} \left(m^2_{d,BB}\right)_{ij} }
\; .
\eea
Note that in the numerics we diagonalize exactly the squark mass
matrices.  After implementing the resummation of the large $\tan\beta$
effects, we use {\it FeynHiggs 2.4.1}~\cite{Heinemeyer:1998yj} to
calculate the Higgs spectrum and the $\rho$ parameter.

In a minimal flavour-violating scenario, all contributions to
$\cseffP$ are suppressed by a factor $m_s/m_b$; hence, large
effects are possible only if some of the mass insertions are
non-vanishing. Let us consider the effect of non-zero mass
insertions in the down sector. The main contribution comes from
penguin diagrams involving a gluino and down squark. From the
approximate expressions given in Ref.~\cite{Lunghi:1999uk} we see
that only $(\delta^d_{RR})_{32}$ and $(\delta^d_{LR})_{32}$
contribute; since the latter is strongly enhanced by a factor
$m_{\tilde g}/m_b$, even modest values of this mass insertion (in
the $10^{-3}$ range) are sufficient to induce large effects on
$\cseffP$. Moreover, since the $m_{\tilde g}/m_b$ enhancement
factor is present exclusively in the contribution to $\cseffP$, a
$(\delta^d_{LR})_{32} \sim O(10^{-3})$ will have a large impact on
$\cseffP$ while being completely irrelevant for any other process.

In the numerical analysis we use the complete formulae for the Wilson
coefficients which can be found in Ref.~\cite{Bobeth:1999ww}. The most
important parameters that we have in this scenario are the gluino mass
$m_{\tilde g}$, the down squark mass eigenvalues and the normalized
insertion $(\delta^d_{LR})_{32}$ .
\begin{figure}[t]
\begin{center}
\epsfig{figure=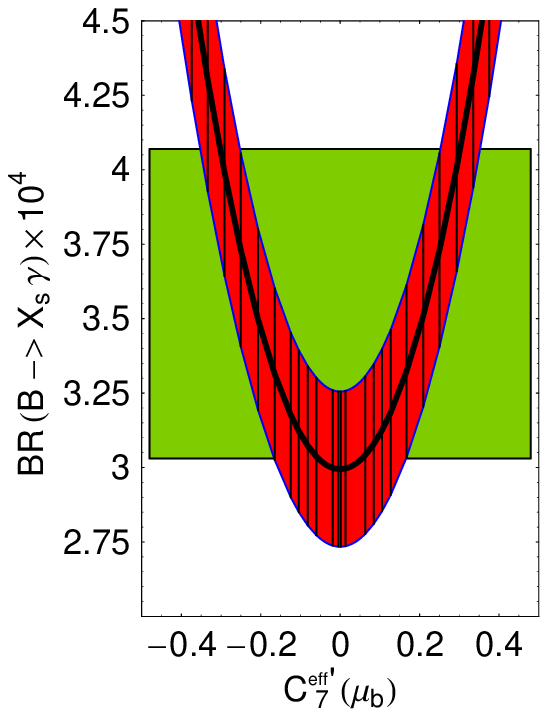,height=0.42\linewidth}
\raisebox{-0.5cm}{\epsfig{figure=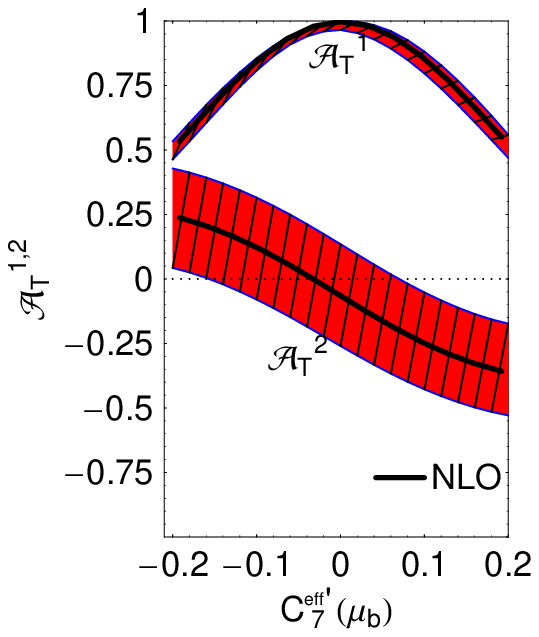}}
\end{center}
\vskip -1cm \caption{Dependence of the $B\to X_s \gamma$ branching
ratio and of the integrated transverse asymmetries ${\cal A}_T^{1,2}$
on the new physics contribution to the Wilson coefficients $\cseffP
(\mu_b)$. The asymmetries are integrated in the low-$s$ region, $1\;
{\rm GeV^2} < s < 6 \; {\rm GeV^2}$. The horizontal band on the left
plot is the 95\% C.L. experimental measurement of ${\cal B}(B\to X_s
\gamma)$. The dashed region on the right plot correspond to adding all
errors in quadrature, including the estimated $\Lambda/m_b$
corrections.}
\label{fig:c7p-int}
\end{figure}
\section{Experimental constraints}
\label{sec:constraints}
The main constraint on our model comes from the $\bar B\to X_s
\gamma$ decay.  The experimental world average for its branching
ratio, with a cut $E_\gamma > 1.6 \; {\rm GeV}$ in the $\bar B$
rest frame, reads~\cite{:2006bi}:
\bea
{\cal B}(B\to X_s \gamma)^{\rm exp}_{E_\gamma>1.6 \; {\rm GeV}}
= \left( 3.55 \pm 0.24^{+0.09}_{-0.10}\pm 0.03 \right) \times 10^{-4} \; ,
\eea
yielding at the 95\% C.L. range $[3.03,4.07] \times 10^{-4}$.
The most recent NNLO results~\cite{Misiak:2006zs,Misiak:2006ab} can be
reproduced by choosing appropriate renormalization scales in the NLO
expressions (see, for instance,
Ref.~\cite{Gambino:2001ew,Hurth:2003dk}). Using the same numerical
inputs as in Ref.~\cite{Misiak:2006ab}, and taking
$(\mu_c,\mu_b,\mu_0) \simeq (1.3,1.4,160)\; {\rm GeV}$, the NLO
central value\footnote{We define NLO according to the analysis
presented in Ref.~\cite{Gambino:2001ew}.} of the branching ratio
coincides with the NNLO one. Other choices of the scales $\mu_{b,c}$
yield the same numerical central value but require to push either one
of the two scales dangerously close to the non-perturbative regime. In
order to implement the estimate of the new class of power corrections
identified in Ref.~\cite{Lee:2006wn} and of the analysis of the photon
energy spectrum presented in Ref.~\cite{Becher:2006pu}, we first
calculated the NLO branching ratio with $E_\gamma > 1 \; {\rm GeV}$,
adopting the above choice of input values and scales, then subtracted
1.65\%, as suggested by the analysis of Ref.~\cite{Lee:2006wn}, and
finally multiplied by the conversion factor~\cite{Becher:2006pu}
\begin{equation}
T =
{\cal B}_{E_\gamma>1.6\; {\rm GeV}}/{\cal B}_{E_\gamma>1\; {\rm GeV}}
=
0.93^{+0.03}_{-0.05}({\rm pert})\pm 0.02({\rm hadr})\pm 0.02({\rm pars}) \; .
\end{equation}
We obtain the following numerical formula in which we allow for
arbitrary new physics contributions to the matching conditions (at the
scale $\mu_0 = 160 \; \gev$) of the leading ($C_{7,8}^{(0)}$ and
$C_{7,8}^{\prime(0)}$) and next-to-leading ($C_{7,8}^{(1)}$) Wilson
coefficients:
\begin{table}
\begin{center}
\begin{tabular}{|l|l|l|l|}
\hline
\spp $a      = 2.98 $ & $a_{7}  = -7.184 + 0.612 \; i$ & $b_{77} = 0.084$ & $b_7 = -0.075$ \\
\spp $a_{77} = 4.743$ & $a_{8}  = -2.225 - 0.557 \; i$ & $b_{88} = 0.007$ & $b_8 = -0.022$ \\
\spp $a_{88} = 0.789$ & $a_{78} =  2.454 - 0.884 \; i$ & $b_{78} = 0.025$ &  \\
\hline
\end{tabular}
\caption{Numerical values of the coefficients that enter Eq.~(\ref{bsgamma}).}
\label{tab:ai}
\end{center}
\end{table}
\bea \label{bsgamma}
\hskip -0.8cm
{\cal B} (\bar B \to X_s \gamma)_{E_\gamma>1.6\; {\rm GeV}}^{\rm th}
& = &
10^{-4}  \Bigg[
a
+ a_{77} \, \left(|\delta C_7^{(0)}|^2 + |\delta C_7^{\prime (0)}|^2\right)
+ a_{88} \, \left(|\delta C_8^{(0)}|^2+ |\delta C_8^{\prime (0)}|^2\right)
\nonumber \\ & & \hskip -4cm
+{\rm Re} \Big(
a_7 \; \delta C_7^{(0)}
+ a_8 \; \delta C_8^{(0)}
+ a_{78} \; \left[ \delta C_7^{(0)} \; \delta C_8^{(0)*}
                   + \delta C_7^{\prime (0)} \; \delta C_8^{\prime(0)*}
            \right] +
b_7 \; \delta C_7^{(1)}
+ b_8 \; \delta C_8^{(1)}
\nonumber \\ & &  \hskip -4cm
+ b_{77} \; \delta C_7^{(0)} \; \delta C_7^{(1)*}
+ b_{88} \; \delta C_8^{(0)} \; \delta C_8^{(1)*}
+ b_{78} \; \left[ \delta C_7^{(0)} \; \delta C_8^{(1)*}
+ \delta C_7^{(1)} \; \delta C_8^{(0)*} \right]
\Big)
\Bigg]\, ,
\eea
where the numbers $a_i$ and $b_i$ are collected in Table~\ref{tab:ai}
and we defined $C_i = C_i^{\rm SM} + \delta C_i$. Eq.~(\ref{bsgamma})
updates the corresponding formula, first presented in
Ref.~\cite{Kagan:1998ym}.
The analyses in Refs.~\cite{Misiak:2006zs,Becher:2006pu} yield ${\cal
B}(B\to X_s \gamma)= (2.98 \pm 0.26) \times 10^{-4}$; we will
therefore assign a theoretical error of 8.7\% to the central values
calculated in Eq.~(\ref{bsgamma}).

In Fig.~\ref{fig:c7p-int} we show graphically the dependence of ${\cal
B}(B\to X_s \gamma)$ on $\cseffP (\mu_b)$. Note that a non-vanishing
$\cseffP (\mu_b)$ can only increase this branching ratio. We also
explore, in a model-independent way, in Fig.~\ref{fig:c7p-int}, the
impact of a non-zero $\cseffP (\mu_b)$ (inside the ${\cal B}(B\to X_s
\gamma)$ allowed range) on the integrated asymmetries ${\cal
A}_T^{1,2}$. It is remarkable how tiny the impact of QCD uncertainties
on the integrated transverse asymmetry ${\cal A}_T^{1}$, even
including an estimated $\Lambda/m_b$ correction of the order of $\pm
10\%$ and a wide soft form factor $\xi_{\perp}(0)$ variation from 0.24
to 0.35 (see Ref.~\cite{Kruger:2005ep}).

The mass insertion $(\delta_{LR}^{d})_{32}$ impacts also the $B_s
-\bar B_s$ mass difference via contributions to the Wilson
coefficients of the pseudo-scalar operators $(\bar s_R b_L)(\bar s_R
b_L)$ and $(\bar s_R^\alpha b_L^\beta)(\bar s_R^\beta b_L^\alpha)$. From 
the 
analysis of Ref.~\cite{Ciuchini:2002uv} it follows that there
are no appreciable contributions to $\Delta m_{B_s}$ for mass
insertions as small as the one we utilize in our numerical
analysis. We checked this statement by explicit calculation of these
contributions.

In the numerics we impose also the constraints from the $\rho$
parameter, Higgs and supersymmetric particle searches: $\delta \rho =
(-0.5 \pm 1.1 ) \times 10^{-3}$, $m_h>89.8\; {\rm GeV}$,
$m_{\chi^\pm}>100\; {\rm GeV}$, $m_{\tilde q}>100\; {\rm GeV}$,
$m_{\chi^0}>40\; {\rm GeV}$.
\begin{figure}
\begin{center}
\epsfig{figure=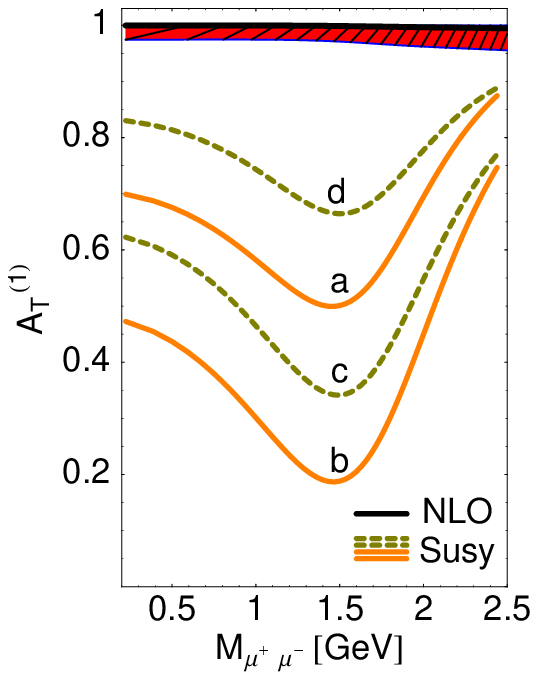,height=0.42\linewidth}
\raisebox{-0.5cm}{\epsfig{figure=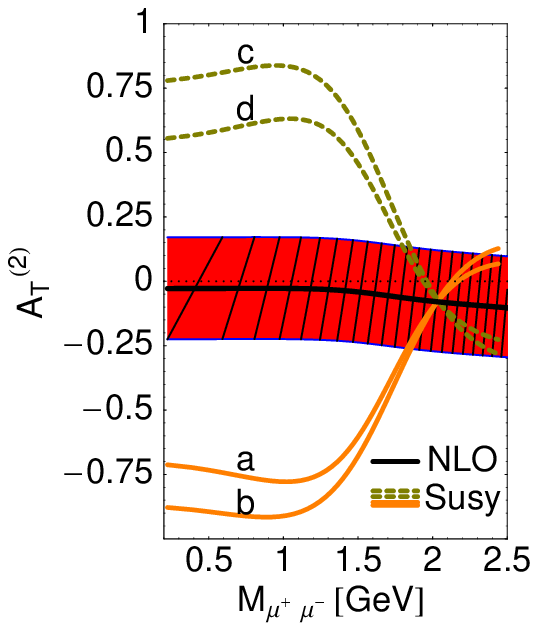}}
\end{center}
\vskip -1.3cm \caption{$A_T^{(1,2)}(s)$ asymmetries versus the
dimuon mass. Thick line correspond to the SM NLO result, while the
band around the thick line is the result of adding all errors in
quadrature. Curves ``a'',``b'',``c'' and ``d'' correspond to
specific choice of parameter space in supersymmetry as explained
in the text.} \label{fig:at12}
\end{figure}
\begin{figure}
\begin{center}
\epsfig{figure=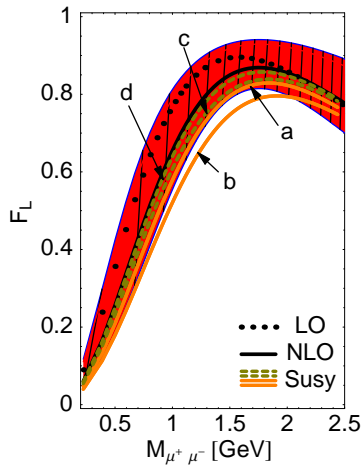,height=.42\linewidth}
\end{center}
\vskip -1cm \caption{$F_L$ versus the dimuon mass. Same
conventions ad in Fig.~\ref{fig:at12}.} \label{fig:alflft}
\end{figure}
\section{Results}
\label{sec:results}
The main results are shown in  Figs.~\ref{fig:at12} and
\ref{fig:alflft}, where the asymmetries $A_T^{(1)}(s)$ and
$A_T^{(2)}(s)$ and the polarization fraction $F_L(s)$ are shown as
a function of the dimuon mass in the SM and in the presence of new
physics  originating from a MSSM, as described in Section 2.

Let us now discuss those three observables in full detail. First,
concerning the SM prediction, and in order to be conservative we have
introduced, in addition to the uncertainties discussed in
\cite{Kruger:2005ep}, a set of extra parameters, one for each spin
amplitude, to explore what the effect of a possible $\Lambda/m_b$
correction could be:
$$ A_{\bot,\|,0}=A_{\bot,\|,0}^0 \left(1+c_{\bot,\|,0} \right) $$
where the `0' superscript stands for the QCD NLO Factorisation
amplitude and $c_{\bot,\|,0}$ are taken to vary in a range $\pm
10\%$. Note that the transversity amplitudes defined in
Eqs.~(\ref{a_perp})--(\ref{a_long}) correspond to physical
transitions; hence the combined effect of any power correction is
described by a single effective parameter for each of these amplitudes
whose size we assumed to be of order $O(10\%)$. It was recently
pointed out~\cite{Grinstein:2004uu} the existence of a class of power
corrections that were taken based on dimensional arguments to be of
order $\frac{C_2}{3 C_7} \frac{\Lambda}{m_b} \sim \frac{\Lambda}{m_b}$
and contribute at leading order to the asymmetries we consider. On the
one hand, we note that these corrections just give a contribution to
the effective parameters that we introduced; on the other one, we
point out that an attempt to estimate these corrections using
light-cone QCD sum rules~\cite{Ball:2006cv,Ball:2006eu} found that
they are indeed suppressed by a factor 12 with respect to the
dimensional arguments. For these reasons we think that the naive
$10\%$ estimate is fair.

Therefore, we allowed these extra parameters to vary independently
in a range of $\pm 10\%$ (i.e. a 20\% range) for each spin
amplitude. The obtained uncertainty was added in quadrature to all
other QCD uncertainties (mainly $m_c/m_b$, scale $\mu$, $f_B$,
$f_K*$, $\lambda_{B,+}$ and $\xi_{\perp}(0)$) and corresponds to
the red region in Figs.~\ref{fig:at12} and \ref{fig:alflft}. It is
clear that while the impact on $A^{(1)}_{\trans}({s})$ is very
small, $A^{(2)}_{\trans}({s})$ and $F_L(s)$ are more affected.
Yet, as explained in the following, while the impact on $F_L(s)$
turns out to be dramatic when distinguishing new physics, it is
not the case for $A^{(2)}_{\trans}({s})$. Notice that the main
source of error in $A^{(1,2)}_{\trans}({s})$ shown in
Figs.~\ref{fig:at12} comes from this extra $\pm 10\% $ uncertainty
and that all other sources are completely negligible as was found
in~\cite{Kruger:2005ep}. This is not the case of $F_{L}(s)$ (see
Fig.~\ref{fig:alflft}), where the size of the other QCD
uncertainties is comparable to this extra uncertainty on the power
corrections.

Concerning new physics that come mainly into play via the Wilson
coefficient $\cseffP$, the fact that $\cseffP$ does not interfere
with $C_7^{\rm eff}$ implies that the experimental constraint on the
new-physics contributions to $\cseffP$ is much looser than the
corresponding one on $C_7^{\rm eff}$. Moreover, it is clear from the
discussion in Sec.~\ref{sec:asymmetries}, that a non-vanishing
$\cseffP$ induces asymmetries already at the LL level. In fact, in
the numerical analysis we find large deviations from the SM only
if $\cseffP$ is non-zero. This make of those asymmetries  a
prominent test of right-handed currents.

In order to see if a specific model, MSSM with R-parity
conservation and non-minimal flavour changing, can lead naturally
to substantial deviations from the SM predictions we have tried to
be as generic as possible. We have explored the space of
parameters of this supersymmetric model in two separate regions
(scenarios A and B) defined by $m_{\tilde g}/m_{\tilde d}$ being
larger (A) and smaller (B) than 1. We have chosen few
representative curves of the two different scenarios, to show
examples of input parameters, but the whole region between those
representative curves and the SM are filled by solutions
consistent with all constraints. In both scenarios we take
$m^2_{u,LL} = m^2_{d,LL} = m^2_{d,RR} = m_{\tilde d}^2 {\mathbf
1}_{3\times 3}$, $m^2_{u,RR} = m_{\tilde u_R}^2 {\mathbf
1}_{3\times 3}$. $\tan\beta = 5$, $\mu=M_1=M_2=M_{H^+}= m_{\tilde
u_R}=1 \; {\rm TeV}$. Note that we choose a low value for
$\tan\beta$; this shows that we do not need to rely on a
large-$\tan \beta$ to see an effect, and ensures automatic
fulfillment of the constraint coming from $B_s \to \mu^+ \mu^-$.

Moreover, we assume that all the entries in $m^2_{u,LR}$ and
$m^2_{d,LR}$ vanish, with the exception of the one that
corresponds to $\left(\delta_{LR}^{d}\right)_{32}$. The remaining
parameters are fixed as follows.
\begin{itemize}
\item Scenario A: $m_{\tilde g} = 1 \; {\rm TeV}$ and $m_{\tilde
d} \in [200,1000] \; {\rm GeV}$. The only non-zero mass insertion
is varied between $-0.1\leq \left(\delta_{LR}^{d}\right)_{32} \leq
0.1$. For each choice of parameters we first check the list of
constraints indicated in the previous section. The curves shown in
Figs.~\ref{fig:at12} and \ref{fig:alflft}, denoted by ``a'' and
``b'', correspond, respectively to $m_{\tilde g}/m_{\tilde
d}=2.5$, $\left(\delta_{LR}^{d}\right)_{32}=0.016$ and $m_{\tilde
g}/m_{\tilde d}=4$, $\left(\delta_{LR}^{d}\right)_{32}=0.036$.
\item Scenario B: $m_{\tilde d} = 1 \; {\rm TeV}$ and $m_{\tilde g}
\in [200,800] \; {\rm GeV}$. The mass insertion is varied in the same
range as Scenario A. The curves shown in Figs.~\ref{fig:at12} and
\ref{fig:alflft} denoted by ``c'' and ``d'', correspond, respectively
to $m_{\tilde g}/m_{\tilde d}=0.7$,
$\left(\delta_{LR}^{d}\right)_{32}=-0.01$ and $m_{\tilde g}/m_{\tilde
d}=0.6$, $\left(\delta_{LR}^{d}\right)_{32}=-0.006$.
\end{itemize}
Interestingly we find that the sign of the asymmetry
$A_T^{(2)}(s)$ for $m_{\mu^+\mu-} < 2 \; {\rm Gev}$ is
anticorrelated with the sign of the mass insertion. The
longitudinal polarization fraction $F_L(s)$, as anticipated, is
poorly sensitive to new physics contributions (already before the
inclusion of the uncertainty from $\Lambda/m_b$ corrections). The
plots in Figs.~\ref{fig:at12} and \ref{fig:alflft} have to be
compared with the corresponding ones in Figs.~3--8 of
Ref.~\cite{Kruger:2005ep}.

\begin{figure}[t]
\begin{center}
\epsfig{figure=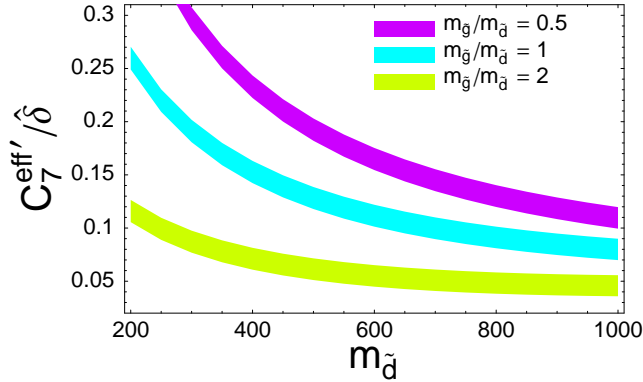,width=.5\linewidth}
\end{center}
\vskip -0.5cm
\caption{Correlation between $\cseffP/{\hat\delta}$
($\hat\delta=(\delta^d_{LR})_{32}/0.005$) and $m_{\tilde d}$ for
various values of $m_{\tilde g}/m_{\tilde d}$.}
\label{fig:c7p-md}
\end{figure}

In Fig.~\ref{fig:c7p-md}, we plot the ratio $\cseffP/ \hat \delta$
as a function of the common down squark mass $m_{\tilde d}$; here
$\hat\delta = (\delta^d_{LR})_{32}/0.005$. The various bands
correspond to the different ratios $m_{\tilde g}/m_{\tilde d} =
(0.5,1,2)$.  Since we use exact diagonalization of the squark mass
matrices, the gluino contribution to $\cseffP$ is not exactly
proportional to the mass insertion and we obtain a band rather
than a line. The combination of Figs.~\ref{fig:c7p-int} and
\ref{fig:c7p-md} allows the immediate translation of a measurement
of $A_T^{(1,2)}(s)$ into information on $m_{\tilde g}$, $m_{\tilde
d}$ and $(\delta^d_{LR})_{32}$ in this framework.

\section{Conclusions}
\label{sec:conclusions} We have shown that the transverse
asymmetries $A_T^{(1,2)}(s)$ are an excellent probe of new physics
induced by right-handed currents. We considered a minimal
supersymmetric model with R-parity conservation and  new
flavour-changing couplings in the right-handed sector. We found
that, after imposing the present experimental constraints,  these
asymmetries still receive huge enhancements and can be visible at
LHCb. The main results are:

\begin{itemize} \item Concerning the SM prediction, we have included an
extra set of parameters to mimic a possible contribution coming
from the subleading $\Lambda/m_b$ correction of order ($\pm
10\%$). We noticed again the robustness of $A_T^{(1)}(s)$ and its
integrated asymmetry ${\cal A}_T^{1}$ compared to $A_T^{(2)}(s)$,
the integrated ${\cal A}_T^{2}$  and $F_L(s)$.

\item Remarkably, already in the low-$\tan \beta$ regime and
taking a very reduced set of parameters ($m_{\tilde g}$,
$m_{\tilde d}$ and $(\delta^d_{LR})_{32}$), sizeable effects are
found in both asymmetries, large enough to disentangle clearly the
supersymmetric contributions of this model from any QCD
uncertainty. Notice, moreover, that $A_T^{(2)}(s)$ provides
information on the sign of the mass insertion. Finally, using the
correlation between $\cseffP/ \hat \delta$ and $m_{\tilde d}$ one
can obtain information on the three free parameters of this model
after a measurement of the asymmetries is done. Concerning the
polarization fraction $F_L(s)$ we did not find large deviations in
any scenario that were not  masked by QCD uncertainties.
\end{itemize}

Negative experimental evidence for deviations in these observables
would result in strong constraints on these flavour-changing
couplings.

\section*{Acknowledgements}
We are grateful to Gudrun Hiller and Mikolaj Misiak for useful
discussions.  Research partly supported by the Department of Energy
under Grant DE-AC02-76CH030000. Fermilab is operated by Universities
Research Association Inc., under contract with the U.S. Department of
Energy. JM acknowledges support from FPA2005-02211, PNL2005-51 and the
Ramon y Cajal Program.

\setlength {\baselineskip}{0.2in}


\begin{thebibliography}{99}

\newcommand{\np}[3]{Nucl. Phys. {\bf B#1} (#2) #3}
\newcommand{\pl}[3]{Phys. Lett. {\bf B#1} (#2) #3}
\newcommand{\pr}[3]{Phys. Rev.  {\bf D#1} (#2) #3}
\newcommand{\prl}[3]{Phys. Rev. Lett. {\bf #1} (#2) #3}
\newcommand{\prp}[3]{Phys. Rept. {\bf #1} (#2) #3}
\newcommand{\ptp}[3]{Prog. Theor. Phys. {\bf #1} (#2) #3}
\newcommand{\zpc}[3]{Z. Phys. {\bf C#1} (#2) #3}
\newcommand{\ibid}[3]{{\it ibid.} {\bf #1} (#2) #3}

\def\euro#1#2#3{{Eur. Phys. J. C} {\bf #1}, #3 (#2)}
\def\eurodirect#1#2#3{{EPJdirect} {\bf C#1}, #3 (#2)}
\def\ijmp#1#2#3{{Int.~J.~Mod.~Phys. A}~{\bf #1}, #3 (#2)}
\def\ib#1#2#3{{\bf#1}, #3 (#2)}
\def\jhep#1#2#3{{J.~High Energy Phys.} {\bf #1}, #3 (#2)}
\def\mpla#1#2#3{{Mod.~Phys.~Lett. A} {\bf #1}, #3 (#2)}
\def\nc#1#2#3{{Nuovo Cimento A}~{\bf#1}, #3 (#2)}
\def\npps#1#2#3{{Nucl.~Phys.~Proc.~Suppl.}~{\bf #1}, #3 (#2)}
\def\prd#1#2#3{{Phys.~Rev. D}~{\bf #1}, #3 (#2)}
\def\rmp#1#2#3{{Rev. Mod. Phys.} {\bf #1}, #3 (#2)}

\bibitem{fs}
Y.~Nir,  arXiv:hep-ph/0109090.
 %%CITATION = HEP-PH 0109090;%%

\bibitem{fs1}
M.~Ciuchini, E.~Franco, A.~Masiero and L.~Silvestrini, J.\ Korean Phys.\ Soc.\  {\bf 45} (2004) S223.
 %%CITATION = HEP-PH 0308013;%%

\bibitem{fs2}
A.~J.~Buras, arXiv:hep-ph/0505175.
 %%CITATION = HEP-PH 0505175;%%

\bibitem{fs3}
R.~Fleischer, arXiv:hep-ph/0505018.
 %%CITATION = HEP-PH 0505018;%%

\bibitem{fs4}
T.~Hurth, AIP Conf.\ Proc.\  {\bf 806}, 164 (2006).
 %%CITATION = APCPC,806,164;%%

\bibitem{dmv1}
T.~Feldmann and T.~Hurth, JHEP {\bf 0411}, 037 (2004).
 %%CITATION = HEP-PH 0408188;%%

\bibitem{dmv2}
S.~Descotes-Genon, J.~Matias and J.~Virto, Phys.\ Rev.\ Lett.\  {\bf 97}, 061801 (2006).
 %%CITATION = HEP-PH 0603239;%%

\bibitem{Melikhov:1998cd}
D.~Melikhov, N.~Nikitin and S.~Simula, Phys.\ Lett.\ B {\bf 442}, 381 (1998).
 %%CITATION = HEP-PH 9807464;%%

\bibitem{CS:etal}
C.~S.~Kim, Y.~G.~Kim, C.-D.~L\"u and T.~Morozumi,
\prd{62}{2000}{034013}.
 %%CITATION = HEP-PH 0001151;%%

\bibitem{CS:etal1}
C.~S.~Kim, Y.~G.~Kim and C.-D.~L\"u, \ibid{64}{2001}{094014}.
 %%CITATION = HEP-PH 0102168;%%

\bibitem{Kruger:2005ep}
F.~Kruger and J.~Matias,  Phys.\ Rev.\ D {\bf 71}, 094009 (2005).
 %%CITATION = HEP-PH 0502060;%%

\bibitem{fm}
T.~Feldmann and J.~Matias, JHEP {\bf 0301} (2003) 074.
 %%CITATION = HEP-PH 0212158;%%

\bibitem{fm1}
J.~Matias, Phys.\ Lett.\ B {\bf 520}, 131 (2001).
  %%CITATION = HEP-PH 0105103;%%

\bibitem{FK:etal}
F.~Kr\"uger, L.~M.~Sehgal, N.~Sinha and R.~Sinha,
\prd{61}{2000}{114028}; \ib{63}{2001}{019901(E)}.
%%CITATION = HEP-PH 9907386;%%

\bibitem{ball}
P.~Ball, J.~M.~Fr\`ere and J.~Matias, Nucl.\ Phys.\ B {\bf 572}, 3
(2000).
 %%CITATION = HEP-PH 9910211;%%

\bibitem{fulvia} P.~Colangelo, F.~De Fazio, R.~Ferrandes and T.~N.~Pham,
  %``Spin effects in rare B --> X/s tau+ tau- and B --> K* tau+ tau- decays in a
  %single universal extra dimension scenario,''
Phys. Rev. D 74, 115006 (2006)
  %%CITATION = HEP-PH 0610044;%%

\bibitem{Aubert:2006vb}
  B.~Aubert {\it et al.}  [BABAR Collaboration],
  %``Measurements of branching fractions, rate asymmetries, and angular
  %distributions in the rare decays B --> K l+ l- and B --> K* l+ l-,''
  Phys.\ Rev.\ D {\bf 73} (2006) 092001.
  %%CITATION = HEP-EX 0604007;%%

\bibitem{Bobeth:1999ww}
C.~Bobeth, M.~Misiak and J.~Urban,  Nucl.\ Phys.\ B {\bf 567}, 153 (2000).
 %%CITATION = HEP-PH 9904413;%%

\bibitem{Lunghi:1999uk}
E.~Lunghi, A.~Masiero, I.~Scimemi and L.~Silvestrini, Nucl.\ Phys.\ B {\bf 568}, 120 (2000).
 %%CITATION = HEP-PH 9906286;%%

 \bibitem{kim}
C.~S.~Kim, Y.~G.~Kim and C.~D.~Lu,
  %``Possible supersymmetric effects on angular distributions in B --> K*  (-->
  %K pi) l+ l- decays,''
  Phys.\ Rev.\ D {\bf 64}, 094014 (2001).
  %%CITATION = HEP-PH 0102168;%%

\bibitem{Misiak:2006zs}
M.~Misiak {\it et al.}, arXiv:hep-ph/0609232.
 %%CITATION = HEP-PH 0609232;%%

\bibitem{Misiak:2006ab}
M.~Misiak and M.~Steinhauser, arXiv:hep-ph/0609241.
 %%CITATION = HEP-PH 0609241;%%

\bibitem{Lee:2006wn}
S.~J.~Lee, M.~Neubert and G.~Paz, arXiv:hep-ph/0609224.
 %%CITATION = HEP-PH 0609224;%%

\bibitem{Becher:2006pu}
T.~Becher and M.~Neubert, arXiv:hep-ph/0610067.
 %%CITATION = HEP-PH 0610067;%%

\bibitem{wilson:coeffs:SM}
A.~J.~Buras and M.~M\"unz, \prd{52}{1995}{186}.
 %%CITATION = HEP-PH 9501281;%%

\bibitem{wilson:coeffs:SM1}
M.~Misiak, \np{393}{1993}{23}; \ib{B439}{1995}{461(E)}.
 %%CITATION = NUPHA,B393,23;%%

\bibitem{wilson:coeffs:SM2}
G.~Buchalla, A.~J.~Buras, and M.~E.~Lautenbacher, \rmp{68}{1996}{1125}.
 %%CITATION = HEP-PH 9512380;%%

\bibitem{wilson:coeffs:SM3}
A.~J.~Buras, in \emph{Probing the Standard Model of
 Particle Interactions}, edited by R.~Gupta \ea\ (Elsevier Science B.V.,
 New York, 1999), p.~281, hep-ph/9806471.
 %%CITATION = HEP-PH 9806471;%%

\bibitem{LRmodel}
P.~Cho and M.~Misiak, \prd{49}{1994}{5894}.
 %%CITATION = HEP-PH 9310332;%%

\bibitem{LRmodel1}
C.~Greub, A.~Ioannisian and D.~Wyler, \pl{346}{1995}{149}.
 %%CITATION = HEP-PH 9408382;%%


\bibitem{Borzumati:1999qt}
F.~Borzumati, C.~Greub, T.~Hurth and D.~Wyler,
\prd{62}{2000}{075005}.
 %%CITATION = HEP-PH 0112126;%%

\bibitem{Borzumati:1999qt1}
T.~Besmer, C.~Greub and T.~Hurth, \np{609}{2001}{359}.
 %%CITATION = HEP-PH 0105292;%%

\bibitem{Borzumati:1999qt2}
L.~Eve\-rett, G.~L.~Kane, S.~Rigolin, L.-T.~Wang, and T.~T.~Wang, \jhep{0201}{2002}{022}.
 %%CITATION = HEP-PH 9911245;%%

\bibitem{pr}
P.~Ball and R.~Zwicky,
  %``B/(d,s) --> rho, omega, K*, Phi decay form factors from light-cone sum
  %rules revisited,''
  Phys.\ Rev.\  D {\bf 71} (2005) 014029
  [arXiv:hep-ph/0412079].
  %%CITATION = PHRVA,D71,014029;%%



\bibitem{Beneke:2001at}
M.~Beneke, T.~Feldmann and D.~Seidel, Nucl.\ Phys.\ B {\bf 612}, 25 (2001).
 %%CITATION = HEP-PH 0106067;%%

\bibitem{Bobeth:1999mk}
C.~Bobeth, M.~Misiak and J.~Urban, Nucl.\ Phys.\ B {\bf 574}, 291 (2000).
 %%CITATION = HEP-PH 9910220;%%

\bibitem{Cho:1993zb}
P.~L.~Cho and M.~Misiak, Phys.\ Rev.\ D {\bf 49}, 5894 (1994).
 %%CITATION = HEP-PH 9310332;%%

\bibitem{Atwood:1997zr}
  D.~Atwood, M.~Gronau and A.~Soni,
  %``Mixing-induced CP asymmetries in radiative B decays in and beyond the
  %standard model,''
  Phys.\ Rev.\ Lett.\  {\bf 79}, 185 (1997)
  [arXiv:hep-ph/9704272].
  %%CITATION = PRLTA,79,185;%%

\bibitem{Atwood2}
D.~Atwood and A.~Soni,
  %``Influence Of Kaonic Resonances On The CP Violation In B $\to$ K* Gamma Like
  %Processes,''
  Z.\ Phys.\  C {\bf 64} (1994) 241
  [arXiv:hep-ph/9401347].
  %%CITATION = ZEPYA,C64,241;%%



\bibitem{Grinstein:2004uu}
  B.~Grinstein, Y.~Grossman, Z.~Ligeti and D.~Pirjol,
  %``The photon polarization in B --> X gamma in the standard model,''
  Phys.\ Rev.\  D {\bf 71}, 011504 (2005)
  [arXiv:hep-ph/0412019].
  %%CITATION = PHRVA,D71,011504;%%

\bibitem{Ulrik} Ulrik Egede, private communication.

\bibitem{exp} B.~Aubert {\it et al.}  [BABAR Collaboration],
  %``Measurement of time-dependent CP-violating asymmetries in $B^0 \to K^{*0}
  %\gamma (K^{*0} \to K^0_S \pi^0)$ decays,''
  Phys.\ Rev.\ Lett.\  {\bf 93} (2004) 201801
  [arXiv:hep-ex/0405082].
  %%CITATION = PRLTA,93,201801;%%



\bibitem{Babu:1999hn}
K.~S.~Babu and C.~F.~Kolda, Phys.\ Rev.\ Lett.\  {\bf 84}, 228 (2000).
 %%CITATION = HEP-PH 9909476;%%

\bibitem{Hamzaoui:1998nu}
C.~Hamzaoui, M.~Pospelov and M.~Toharia, Phys.\ Rev.\ D {\bf 59}, 095005 (1999).
 %%CITATION = HEP-PH 9807350;%%

\bibitem{Isidori:2001fv}
G.~Isidori and A.~Retico, JHEP {\bf 0111}, 001 (2001).
 %%CITATION = HEP-PH 0110121;%%

\bibitem{Buras:2002wq}
A.~J.~Buras, P.~H.~Chankowski, J.~Rosiek and L.~Slawianowska, Phys.\ Lett.\ B {\bf 546}, 96 (2002).
 %%CITATION = HEP-PH 0207241;%%

\bibitem{Buras:2002vd}
A.~J.~Buras, P.~H.~Chankowski, J.~Rosiek and L.~Slawianowska, Nucl.\ Phys.\ B {\bf 659}, 3 (2003).
 %%CITATION = HEP-PH 0210145;%%

\bibitem{Dedes:2002er}
A.~Dedes and A.~Pilaftsis, Phys.\ Rev.\ D {\bf 67}, 015012 (2003)
 %%CITATION = HEP-PH 0209306;%%

\bibitem{Foster:2005wb}
J.~Foster, K.~i.~Okumura and L.~Roszkowski, JHEP {\bf 0508}, 094 (2005).
 %%CITATION = HEP-PH 0506146;%%

\bibitem{Heinemeyer:1998yj}
S.~Heinemeyer, W.~Hollik and G.~Weiglein, Comput.\ Phys.\ Commun.\  {\bf 124}, 76 (2000).
 %%CITATION = HEP-PH 9812320;%%

\bibitem{:2006bi}
Heavy Flavor Averaging Group (HFAG), arXiv:hep-ex/0603003.
 %%CITATION = HEP-EX 0603003;%%

\bibitem{Gambino:2001ew}
  P.~Gambino and M.~Misiak,
  %``Quark mass effects in anti-B --> X/s gamma,''
  Nucl.\ Phys.\ B {\bf 611}, 338 (2001).
%  [arXiv:hep-ph/0104034].
  %%CITATION = HEP-PH 0104034;%%

\bibitem{Hurth:2003dk}
T.~Hurth, E.~Lunghi and W.~Porod, Nucl.\ Phys.\ B {\bf 704}, 56 (2005).
 %%CITATION = HEP-PH 0312260;%%

\bibitem{Kagan:1998ym}
  A.~L.~Kagan and M.~Neubert,
  %``{QCD} anatomy of B --> X/s gamma decays,''
  Eur.\ Phys.\ J.\ C {\bf 7}, 5 (1999)
  [arXiv:hep-ph/9805303].
  %%CITATION = HEP-PH 9805303;%%

\bibitem{Ciuchini:2002uv}
  M.~Ciuchini, E.~Franco, A.~Masiero and L.~Silvestrini,
  %``b --> s transitions: A new frontier for indirect SUSY searches,''
  Phys.\ Rev.\ D {\bf 67}, 075016 (2003).
  [Erratum-ibid.\ D {\bf 68}, 079901 (2003)]
%  [arXiv:hep-ph/0212397].
  %%CITATION = HEP-PH 0212397;%%

\bibitem{Ball:2006cv}
  P.~Ball and R.~Zwicky,
  %``Time-dependent CP asymmetry in B --> K* gamma as a (quasi) null test of
  %the standard model,''
  Phys.\ Lett.\  B {\bf 642}, 478 (2006)
  [arXiv:hep-ph/0609037].
  %%CITATION = PHLTA,B642,478;%%

\bibitem{Ball:2006eu}
  P.~Ball, G.~W.~Jones and R.~Zwicky,
  %``B --> V gamma beyond QCD factorisation,''
  Phys.\ Rev.\  D {\bf 75}, 054004 (2007)
  [arXiv:hep-ph/0612081].
  %%CITATION = PHRVA,D75,054004;%%


\end{thebibliography}
\end{document}